\documentclass[conference,letterpaper,10pt]{IEEEtran}
\ifCLASSINFOpdf
\else
\fi
\ifCLASSOPTIONcompsoc
 \usepackage[caption=false,font=normalsize,labelfont=sf,textfont=sf]{subfig}
\else
\usepackage[caption=false,font=footnotesize]{subfig}
\fi

\usepackage[english]{babel}
\usepackage[utf8x]{inputenc}
\usepackage[T1]{fontenc}

\usepackage{amsmath}
\usepackage{graphicx}
\usepackage[colorinlistoftodos]{todonotes}
\usepackage[colorlinks=true, allcolors=blue]{hyperref}
\usepackage{algorithm}
\usepackage[noend]{algpseudocode}

\begin{document}
%
\title{HeNet: A Deep Learning Approach on Intel\textsuperscript{\textregistered} Processor Trace for Effective Exploit Detection}

\author{\IEEEauthorblockN{Li Chen}
\IEEEauthorblockA{
Security and Privacy Lab, Intel Labs\\
Hillsboro, OR 97124}
\and
\IEEEauthorblockN{Salmin Sultana}
\IEEEauthorblockA{
Security and Privacy Lab, Intel Labs\\
Hillsboro, OR 97124}
\and
\IEEEauthorblockN{Ravi Sahita}
\IEEEauthorblockA{
Security and Privacy Lab, Intel Labs\\
Hillsboro, OR 97124}
}


%


\maketitle
\begin{abstract}


This paper presents HeNet, a hierarchical ensemble neural network, applied to classify hardware-generated control flow traces for malware detection. Deep learning-based malware detection has so far focused on analyzing executable files and runtime API calls. Static code analysis approaches face challenges due to obfuscated code and adversarial perturbations. Behavioral data collected during execution is more difficult to obfuscate but recent research has shown successful attacks against API call based malware classifiers. We investigate 
control flow based characterization of a program execution to build robust deep learning malware classifiers. 

HeNet consists of a low-level behavior model and a top-level ensemble model. The low-level model is a per-application behavior model, trained via transfer learning on a time-series of images generated from control flow trace of an execution. We use Intel\textsuperscript{\textregistered} Processor Trace enabled processor for low overhead execution tracing and design a lightweight image conversion and segmentation of the control flow trace. The top-level ensemble model aggregates the
behavior classification of all the trace segments and detects an attack. The use of hardware trace adds portability to our system and the use of deep learning eliminates the manual effort of feature engineering. 

We evaluate HeNet against real-world exploitations of PDF readers. HeNet achieves 100\% accuracy and 0\% false positive on test set, and higher classification accuracy compared to classical machine learning algorithms.
\end{abstract}


%
\IEEEpeerreviewmaketitle

\section{Introduction}
According to recent security threat reports, one million malware variants hit the internet every day while new malware samples found in Q3 2017 increased to a record high of 57.6 million~\cite{Behrends17}\cite{Minihane17}. 
Traditional signature-based malware detection techniques are growing ineffective as the malware authors use code obfuscation, polymorphism, etc. to automate writing new malware variants and zero-day exploits to create new malware. Hence, the security community are pursuing machine learning-based detection approaches to adapt to the rapidly changing malware ecosystem. Neural networks able to automatically learn features have been proven to detect and classify complex 
malware with high accuracy. While significant progress in deep learning makes it promising for automated malware analysis, further work is needed to build \textit{robust} and \textit{scalable} malware classifiers.

Existing deep learning-based malware detection and classification systems learn features from static analysis of an executable file or runtime behavioral analysis. The static analysis approaches build neural networks using header, instruction opcodes, or raw bytes of an executable file and classify the file before execution~\cite{RaffSylvester2017}\cite{Davis2015}\cite{Raff2017}. The dynamic analysis approaches model user or kernel API call sequences over time for a program execution~\cite{Dahl2013}\cite{Saxe2015}\cite{Kim2016}\cite{Huang2016}\cite{Kolosnjaji2016}. Despite their potential, the neural network malware classifiers are susceptible to \textit{adversarial inputs}. By crafting adversarial malware samples~\cite{Grosse2016} or API call sequences~\cite{Hu2017}\cite{Rosenberg2017}, an  attacker can mislead the malware classifier and achieve high misclassification rate without affecting the malware functionality. Generally, it is a lot more challenging to perturb dynamically gathered features. The API call-based dynamic features, however, have been shown to be vulnerable to black-box attacks. The proposed attacks are effective against many classifiers including recurrent neural network variants, feed forward deep neural networks, and traditional machine learning classifiers. To be noted, API call-based malware analysis requires significant manual feature engineering to select the API calls most relevant to malware execution.

We investigate more sophisticated dynamic features to build a robust deep learning malware classifier. In this paper, we describe a malware detection system powered by HeNet, a hierarchical ensemble neural network, that classifies the control flow trace of program execution. Control flow analysis has been widely used in security to mitigate memory corruption attacks by enforcing pre-specified security policies over a program execution~\cite{Abadi09}. HeNet, however, is the first malware detection approach to apply deep learning on fine-grained control flow traces.  We build a per-application deep-learning behavioral model to distinguish benign and malicious executions of the application. The key observation underlying our work is that malware exploits temporarily or persistently execute malicious code in the context of a victim process and hence, change the original control flow of the application. 

We utilize Intel\textsuperscript{\textregistered} Processor Trace (Intel\textsuperscript{\textregistered} PT), a low overhead tracing capability in the CPU, to get the complete control flow audit of a program execution. Intel\textsuperscript{\textregistered} PT records the non-deterministic control flow transfers, contextual, timing, etc. information and transmits the data in highly compressed trace packets. The control flow packets combined with the program binary can be used to reconstruct the exact sequence of instructions executed but the post-processing takes a long time due to the high volume of trace data. We, however, notice that the control flow packets can be used as an encoded representation of the control flow transfers over a program execution. We convert the payload of control flow packets to a stream of pixels and represent the control flow as a time series of images. Utilizing low level trace packets eliminates manual feature engineering and makes our system portable across different operating systems and hardware.

We train a deep neural network on the sequence of control flow images to build an  application behavioral model. To accelerate the training on a large number of images, we use deep transfer learning~\cite{pan2010survey}\cite{yosinski2014transferable}. Each trace is represented by a time series of images whose label is classified by the low-level model. An ensemble model is imposed on top to aggregate the behavior classifications over the entire trace.

The contributions of this paper are summarized as follows:
(1)  We develop the first (to the best of our knowledge) deep learning approach to malware detection using fine-grained control flow traces of an application execution. We present the architecture of HeNet, a hierarchical ensemble neural network, that classifies a control flow trace to detect malware.

(2) We utilize Intel\textsuperscript{\textregistered} PT, a low overhead tracing hardware in commodity processors, to collect control flow trace. 
 The low level trace based features make HeNet extensible and portable to various systems.

(3) We represent the control flow trace as a time series of images and define behavioral anomaly detection as an image classification problem. We apply transfer learning to achieve high accuracy and faster training on large number of trace images.

(4) We evaluate HeNet against real-world exploitations of Adobe\textsuperscript{\textregistered} Reader 9.3 on Windows\textsuperscript{\textregistered} 7 32 bit. The experimental results show 100\% accuracy to classify malicious and benign PDFs and 0\% false positive. We show that HeNet produces much higher classification accuracy and lower false positive rate compared to classical machine learning algorithms.
 
The rest of the paper is organized as follows: Sec.~\ref{sec:background} discusses the threat model and a brief background on Intel\textsuperscript{\textregistered} PT. Sec.~\ref{sec:overview} presents the design overview of our anti-malware system. Sec.~\ref{sec:henet} presents HeNet architecture in detail. HeNet performance results are reported in Sec.~\ref{sec:exp}. We discuss the related work in Sec.~\ref{sec:related} and conclude in Sec.~\ref{sec:conclusion}.
\section{Background and Threat Model}
\label{sec:background}
\subsection{Threat Model}
We intend to detect malware attacks by classifying control flow transfers over an application execution. Hence, we consider the attacks that change the execution flow of a benign application. This threat model covers a wide range of malware attacks, including control flow attacks such as Return-Oriented-Programming (ROP), Jump-Oriented-Programming (JOP) attacks, fileless malware that run under the cover of benign applications, zero-day attacks. 

HeNet learns the control flow model of an application execution from the sequence of control flow packets produced by Intel\textsuperscript{\textregistered} PT. Hence, we assume that the Intel\textsuperscript{\textregistered} PT hardware is trusted so that the generated trace is complete and accurate. We also assume that the integrity of trace is protected in memory and while in use for decoding, training, and classification. 

\subsection{ Intel\textsuperscript{\textregistered} Processor Trace (Intel\textsuperscript{\textregistered} PT)}

Intel\textsuperscript{\textregistered} PT is a low overhead debugging hardware that captures the \textbf{complete} execution  trace of monitored applications. It captures information about application execution on each hardware thread using dedicated hardware so that the trace can be post-processed to reconstruct exact program flow. This capability is particularly useful for baremetal malware analysis, targeted to track evasive malware that can otherwise bypass software instrumentation based malware detection.  
 
Intel\textsuperscript{\textregistered} PT captures control flow, timing, and other contextual information in highly compressed trace packets. 
We utilize the Target IP (TIP) and Taken Not-Taken (TNT) packets to characterize an application execution. TIP packets record the target address of indirect branches, exceptions, interrupts. TNT packets track the direction of conditional branches within basic blocks. A TNT packet contains a number of conditional branches, where each bit corresponds to a conditional branch. If a conditional branch is taken during execution, a bit of value 1 is recorded; 0 otherwise. 
\section{Malware Detection System Overview}
\label{sec:overview}
\begin{figure}[h!]
	\centering
    \vspace{-0.2in}
  \includegraphics[width=0.65\linewidth]{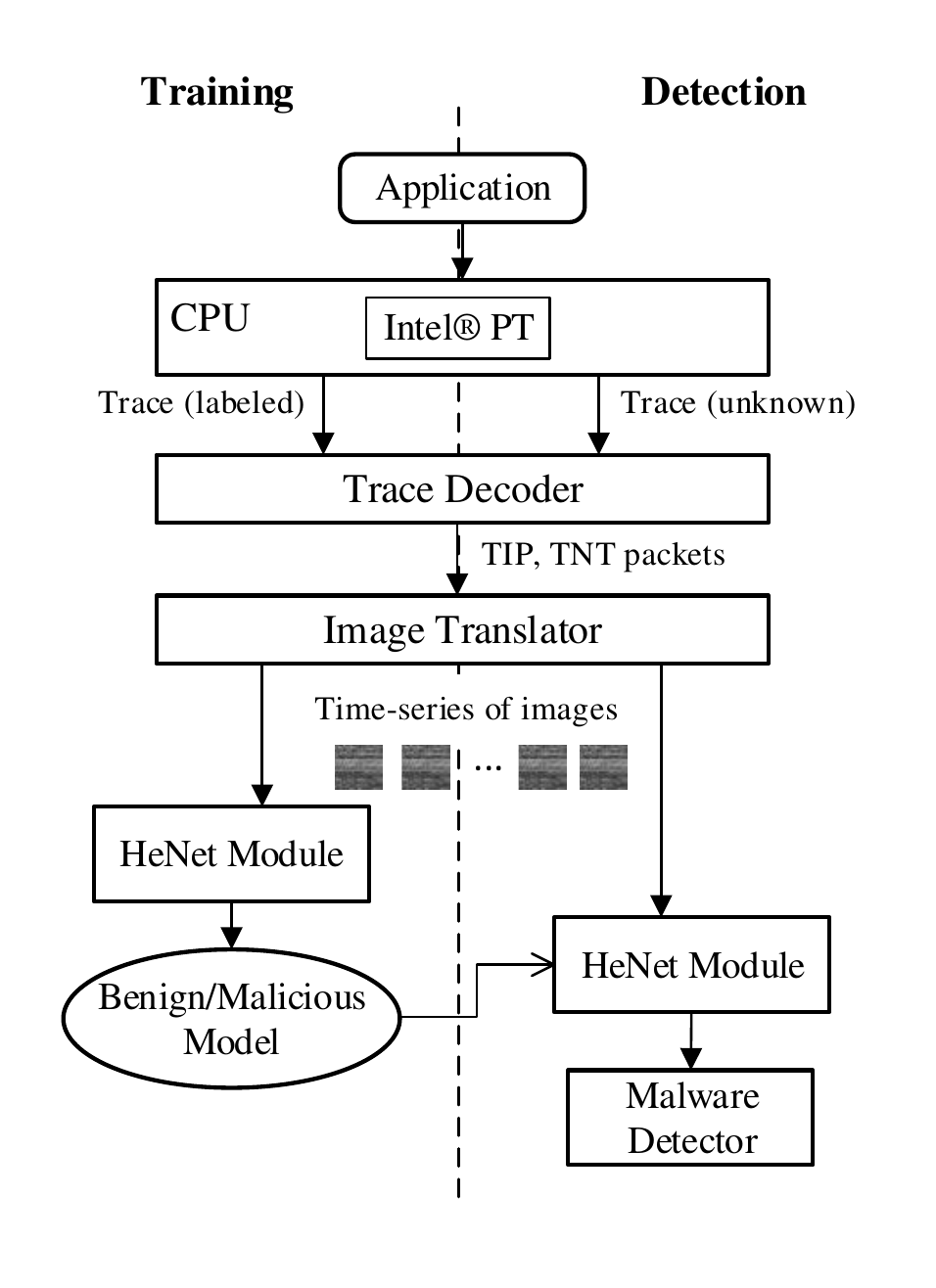}
    \vspace{-0.2in}
  \caption{System overview of malware detection using Intel\textsuperscript{\textregistered} PT and HeNet}
  \label{fig:overview}
\end{figure}

Figure~\ref{fig:overview} shows the system overview of malware detection using HeNet on Intel\textsuperscript{\textregistered} PT trace. We train HeNet, consisting of a low-level behavior and top-level ensemble model, to build a benign/malicious classification model of an application execution. During testing, we apply the HeNet behavior model to detect anomalous trace/execution segments and its ensemble model to decide on an attack over the full trace.
We configure Intel\textsuperscript{\textregistered} PT to monitor an application at runtime and generate execution trace. We use both benign and malicious traces of an application to train HeNet. We refer to an execution trace as \textit{benign} when the benign version of an application is executed and as \textit{malicious} when the application is exploited to conduct an attack. Given a raw PT trace, the \textit{Trace Decoder} decodes the trace packets and extracts the sequence of TIP and TNT packets. A data pre-processing component, \textit{Image Translator}, converts the TNT and TIP packets to a 1-dimensional array of pixels. Due to the high volume of trace data, it is computationally infeasible to build a neural network on the whole pixel array. Instead, we segment the pixel array into $n$ sub-arrays and establish a low-level behavior model as a part of the hierarchical structure of HeNet. Thus, the \textit{Image Translator} converts the execution trace to a time series of images. The \textit{HeNet Module} performs transfer learning on the sequence of trace images to learn a benign/malicious behavior model. 

At testing, we perform segmentation and image conversion of a given trace, as similar to training. The \textit{HeNet Module} applies the behavior model on each image to classify the execution behavior in that segment, and uses top-level ensemble to aggregate the behaviors learned from all the segments. The aggregated probability is fed into a \textit{Malware Detector} which detects an attack if the probability exceeds a threshold. 
\section{Control Flow to Characterize Execution} \label{sec:preprocess}
We utilize the TIP and TNT packets in Intel\textsuperscript{\textregistered} PT trace to characterize an execution. We design a lightweight image conversion technique that converts the payload of a TIP/TNT packet to a stream of pixels. A pixel value is [0, 255]. 
\subsection{Taken Not-Taken (TNT) packet conversion}
There are two variants of TNT packets: (1) Short TNT packet of 1 byte, (2) Long TNT packet of up to 8 bytes. The short TNT packet has the least significant bit as 1-bit header and may contain from 1 to 6 TNT bits. The long TNT packet has the least significant 2 bytes as 2-byte header and may contain up to 47 TNT bits. 
The last valid TNT bit is followed by a trailing 1, or \textit{stop-bit}. If the TNT packet is not full, 
the \textit{stop-bit} moves up and the trailing bits 
are filled with 0's.

We convert the short TNT packet to an unsigned 1-byte value representing 1 pixel intensity. For long TNT packets, we remove the headers and find the length of the packet payload. We convert each byte of payload to a pixel value. Hence, a long TNT packet may generate from 1 to 6 pixels.

\subsection{Target IP (TIP) packet conversion}
The TIP packet has 1 byte header, where the most significant 3 bits specify how the target address of an indirct branch is compressed in payload. Due to compression, the size of target IP payload may vary from 0 to 8 bytes.

We remove the TIP packet header and extract the target IP payload. 
Using the IP payload and last saved IP, we reconstruct the full target address of the indirect branch. We use auxiliary runtime binary load information to find the binary which the target address belongs to. We identify each binary with a 1 byte ID. To normalize the impact of ASLR, we calculate the offset of the target address from the binary base address. We represent the target address as a <binary ID, offset> pair and convert the 1 byte ID and 4 byte offset to an array of 5 pixels.


\section{HeNet: Hierarchical Ensemble Neural Network} \label{sec:henet}
In this section, we present the detailed architecture of HeNet, designed for Intel\textsuperscript{\textregistered} PT trace. HeNet consists of a low-level deep learning model to classify benign/malicious behavior of trace segments, and a high-level ensemble model to aggregate behaviors learned from all the trace segments. 



\subsection{Segmentation and Image Conversion} \label{subsec:seg}
After the pixel conversion of TNT and TIP packets as described in Section \ref{sec:preprocess}, an execution trace is represented as a one-dimensional pixel array. We denote the pixel array by $X \in [0,255]^L$, i.e., $X$ has length $L$ and each element is within 0-255. Next, we divide $X$ into $n$ segments, where each trace segment is of length $m^2$. If $L$ is not a multiple of $m^2$, we pad the last segment with 0's. 

Each segment is mapped to a two-dimensional $m\times m$ array and represented as a gray-scale image. Thus, $X$, the pixel representation of a trace, is transformed to a time series of images $(I_1, I_2, \cdots, I_n)$, where each $I_k \in [0, 255]^{m \times m}$.  Note that the sequence-to-image transformation only resizes the data shape and no information is lost from resizing.


\subsection{Low-Level Behavior Model}
The low-level model in HeNet is a Deep Neural Network (DNN) to classify the behavior of a trace segment as benign or malicious. As we use the image representeation of a trace segment, we apply deep learning to extract the textural features of the trace images. Visual similarities among the benign trace segment images and among the malicious trace segment images motivate us to frame the problem as an image classification problem. 


\subsubsection{Training Dataset}
The time series of trace images are the training dataset for the low-level DNN model. We utilize the original label of the Intel\textsuperscript{\textregistered} PT trace to label the sequence of images. Noisy ground truth, however, imposes a challenge here. 
In supervised learning, the labels of training data are assumed to be absolutely correct. Our ground truth is noisy, since a malicious trace may contain benign execution until the attack happens. Hence, if a trace $X^i$ is malicious, we first identify the time of attack $k$ and then label $(I_k^i, I_{k+1}^i, \cdots, I_n^i)$ as malicious. For a benign trace $X^i$, all the segmented images $(I_1^i, I_2^i, \cdots, I_n^i)$ are labeled as benign.

\subsubsection{Training via Transfer Learning}
Intel\textsuperscript{\textregistered} PT generates a large volume of data and segmentation makes the low-level sample size even larger. Hence, training a neural network from scratch may not be the optimal solution. We perform transfer learning~\cite{pan2010survey} to train the behavior model. 
We transfer the knowledge learned from the state-of-the-art DNNs on large image sets and retrain the layers suitable for our classification. 

Currently, HeNet framework supports transfer learning using a variety of networks including Inception~\cite{szegedy2015going}\cite{szegedy2016rethinking}\cite{szegedy2017inception} and VGG~\cite{simonyan2014very} networks, pre-trained on ImageNet datasets. 
The resizing parameter $m$ is set to an appropriate value. The channels of gray-scale images are augmented by replicating the gray channel into RGB channels. Our empirical analysis shows that the choice of 
a particular transfer learning network does not make significant difference in classification performance. HeNet, however, shows higher accuracy and lower false positive rates compared to classical machine learning algorithms and neural networks trained from scratch. Transfer learning also helps HeNet achieve faster model convergence (within 10 epochs in our experiments). 
\subsection{Top-level Ensemble Model}
Given an execution trace, represented as image sequence $(I_1, I_2, ..., I_n)$, the low-level model produces corresponding behavioral probabilities $(p_1, p_2, ..., p_n)$. The ensemble model aggregates the probabilities by computing the average, $\overline{p} =\frac{1}{n}\sum_{i=1}^{n} p_i$. We compare $\overline{p}$ with a pre-specified threshold to decide whether the execution corresponds to an attack or not. 


HeNet is summarized in Algorithm~\ref{alg:henet}.
\begin{algorithm}[H]\label{alg:henet}
\caption{HeNet on Intel\textsuperscript{\textregistered} PT}\label{alg:henet}
\begin{algorithmic}[1]
\State \textbf{Goal:} Malware detection.
\State \textbf{Input:} Intel\textsuperscript{\textregistered} PT trace, resizing parameter $m$. 
\State \textbf{Step 1} Convert TNT and TIP packets to pixels of value [$0$, 255]. Denote pixel array by $X$.  
\State \textbf{Step 2} Slice $X$ into $n$ segments, where $n = \frac{L}{m^2}$. $L$ is the length of $X$ and a segment is of length $m^2$. 
\State \textbf{Step 3} Convert each segment to an image of $m \times m$. Denote an image by $I_k$ with $k \in \{1,2,\cdots, n\}$.
\State \textbf{Step 4} Apply low-level deep learning behavior model on $I_k$ and obtain probability of each segment as $p_k$.
\State \textbf{Step 5} Predict the label of $X$ by averaging $(p_1, ...,p_n)$.
\end{algorithmic}
\end{algorithm}
\section{Performance Evaluation}\label{sec:exp}
In this section, we present the performance results of HeNet in detecting real-world attacks. We evaluate HeNet with respect to malware detection accuracy and false positive rate. We compare the performance of HeNet low-level model with a neural network trained from scratch and three classical machine learning algorithms. 
\subsection{Experimental Setup}
To show the effectiveness of our malware detection system, we use ROP attacks against Adobe\textsuperscript{\textregistered} Reader 9.3 in Windows\textsuperscript{\textregistered} 7 32 bit. In our experiments, we use a set of 348 benign and 299 malicious PDF samples. 
We utilize Intel\textsuperscript{\textregistered} VTune\textsuperscript{TM} Amplifier driver to collect Intel\textsuperscript{\textregistered} PT trace and develop a  decoder to extract TNT and TIP packets. Out of total 647 traces, 53 are set aside for out-of-sample testing. From the remaining traces, we choose train-validation-test ratio as $0.8:0.1:0.1$. 
We set the resizing parameter $m$ to 224 and the gray-scale images are of size $224\times 224$. Thus, the training and validation sets contain 419075 benign and 276110 malicious trace images and the test set have 48886 benign and 29028 malicious images. 


\subsection{Experimental Results}
\subsubsection{Low-level behavior model performance}
\begin{figure}[t!]
\centering
    \subfloat[ROC of behavior model. The area under curve is 0.989]{
    	\includegraphics[width=0.43\columnwidth]{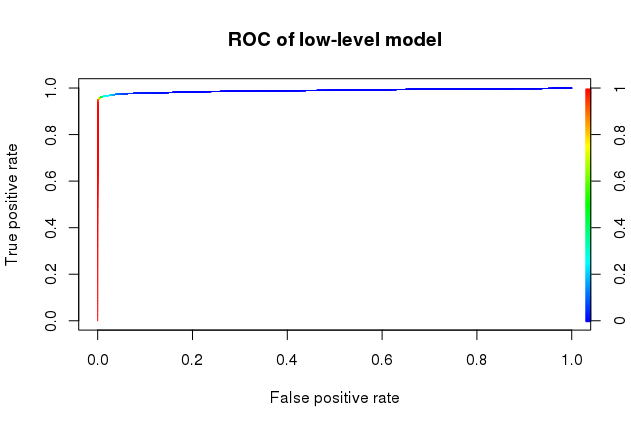} \label{fig:ROC}
	}
	\hfill	
	\subfloat[Plot of confusion matrix. 
    Accuracy is 0.981 and false positive rate is 0.0073.]{   
    	\includegraphics[width=0.5\columnwidth]{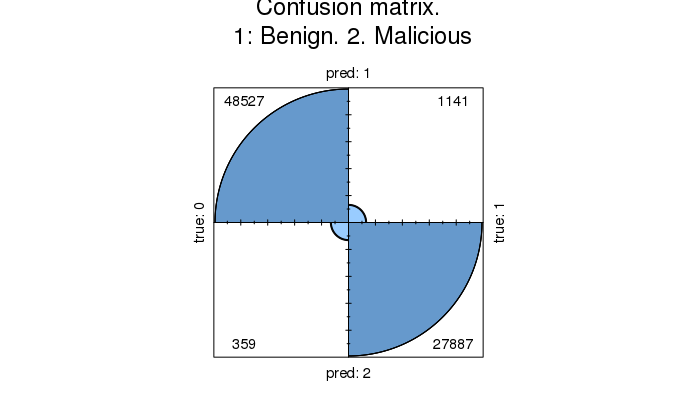} \label{fig:confusion}
    }
    \caption{Performance of low-level behvaior model}
    \label{fig:lowmodel}
	\end{figure}

To train the low-level model, we utilize Inception-BN at iteration 126 as the pre-trained model for transfer learning with learning rate at 0.05 and batch size at 100. We retrain the last fully connected layer and the softmax layer on trace segment images. Figure~\ref{fig:ROC} shows the ROC curve of the low-level behavior model. The area under the curve is 0.989. Figure~\ref{fig:confusion} shows that the behavior model achieves 0.981 accuracy and 0.0073 false positive rate on the test set. Thus, HeNet achieves low false positive rate without sacrificing accuracy. Utilizing transfer learning, the training converges within 10 epochs. 




\begin{figure}[th]
\centering
\includegraphics[scale=0.43]{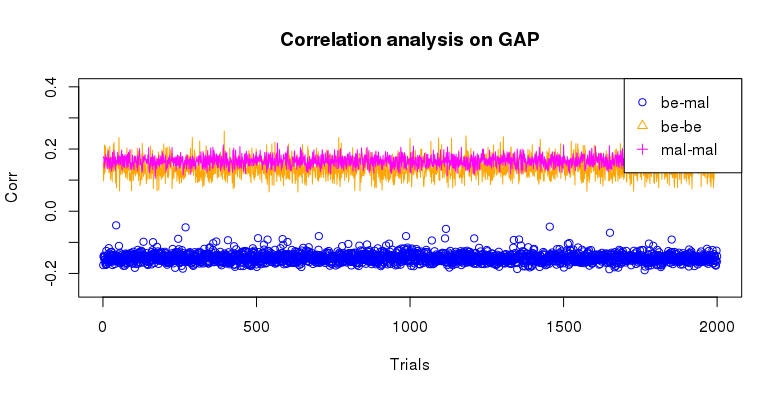}
\includegraphics[height=2.5in, width=3.3in]{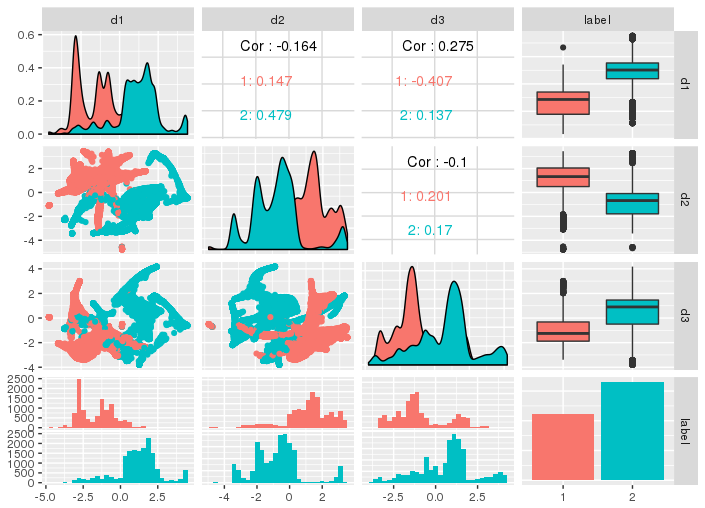}
\caption{Visualization of the features learned from HeNet low-level model.}
\label{fig:feature}
\end{figure}

Next, we examine the features extracted from the low-level behavior model using transfer learning. We randomly select 22\% of benign and 55\% of malicious behavior images from test set, feed them to the trained low-level model, and compute the global average pooling. Then we apply t-distributed stochastic neighbor embedding (t-SNE)~\cite{maaten2008visualizing} on the global average pooling and obtain the first three dimensions. 
Figure~\ref{fig:feature} shows the correlation plot (top), t-SNE scatter plots (lower-triangular), t-SNE density plots (diagonal), correlation (upper triangular), and label box-plots (upper triangular). As we see, the first three t-SNE dimensions are well separated. The correlation among the malicious trace images and correlation among the benign trace images are positive. The correlation among the malicious and benign trace images is negative. 


We present the performance of the low-level model when trained via Inception transfer learning  vs. trained from scratch. We note that training a neural network from scratch takes significantly long time. Hence, we resize the trace images to low-resolution images of size $28 \times 28$ with gray-scale channel and select LeNet~\cite{lecun1998gradient} to train from scratch. Since LeNet architecture is much shallower,  training per epoch does not take too long, yet the model converges much slower. 

We also compare the performance of HeNet with classical machine learning algorithms including random forest, naive Bayes, and $k$-nearest neighbor on high and low resolution vectorized images. 
For this comparison, we sub-sample $17\%$ of the training data for 5 times and compute the average accuracy and false positive. For $224\times224$ images, we apply principal component analysis to reduce dimension from 50176 to 1000. 

\begin{table} 
\centering
  \label{tab:comparison}
  \caption{Comparison of HeNet with other classifiers}
  \begin{tabular}{||l|l|l|l||}
  \hline
    Behavior model &  Accuracy & FPR & Resolution \\ \hline  
     HeNet low-level model & 0.98 & 0.0073 & 224$\times$224$\times$3 \\
     Low-resolution model & 0.90   &  0.1& 28$\times$28$\times$1\\
     3-nearest neighbor + PCA & 0.80 &  0.05& $224^2 \rightarrow 1000$ \\
     Random forest + PCA & 0.63 & 0.02& $224^2 \rightarrow 1000$ \\
     3-nearest neighbor & 0.75 &  0.4 & $28^2$\\
     Naive Bayes & 0.82 & 0.23& $28^2$\\ \hline
\end{tabular}
 \label{tab:comparison}
 \vspace{-0.2in}
\end{table}

Table~\ref{tab:comparison} presents the comparative analysis of HeNet low-level model, LeNet low-resolution model, nearest neighbor, and random forest on both high and low resolution images. Among all the algorithms, the HeNet low-level model achieves the highest behavior classification accuracy and lowest false positive rate. The low-resolution model is trained on $28\times28$ grey-scale images, resized from the $224\times224$ images. The performance degradation of the low-resolution model may be due to much shallower architecture compared to Inception-v1, fewer image channels, or information loss from image resizing. The low-resolution model, however, may provide a performance baseline for the deep learning image classifiers. We further investigate the low-resolution model with deeper architecture on three-channel images in Sec \ref{sec:conclusion}. We expect the machine learning algorithms will have performance improvement if more domain expertise is leveraged for feature extraction. On the other hand, it highlights the advantage of deep learning in greatly saving the cost of feature engineering. 



\subsubsection{Top-level ensemble model performance}
The ensemble model aggregates the low level execution behaviors by computing an average of the behavior probabilities. To determine whether an execution trace is benign or malicious, we set a threshold of 0.5. On our test set, we obtain 100\% detection accuracy and 0 false positive rate. 

\begin{figure}
\centering
\includegraphics[scale=0.48]{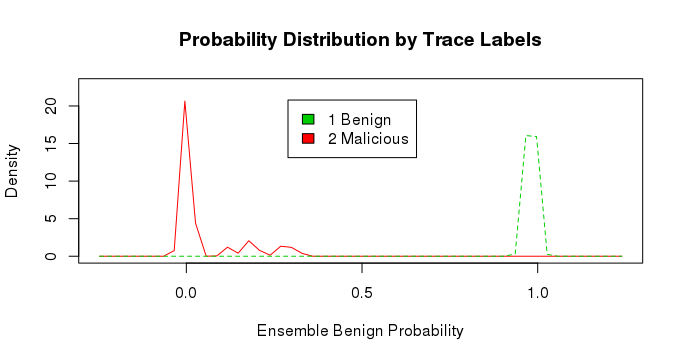}
\vspace{-0.1in}
\caption{Density of the ensemble benign probabilities. HeNet produces high confidence predictions.}
\label{fig:density}
\vspace{-0.1in}
\end{figure}

The ensemble probabilities of being benign for the 31 benign test traces have a mean of 0.98 and a minimum of 0.96. The $p$-value of a two-sided Wilcoxon rank sum test on the ensemble probabilities of being benign for the 31 benign traces and 
22 malicious traces is $4.3\times 10^{-15}$, indicating statistically distinguishable prediction scores. 
Figure~\ref{fig:density} shows the density of ensemble benign probability across benign and malicious test cases. Thus, HeNet achieves high confidence classification at both low-level behavior model and top-level ensemble model.

\subsection{Time Series of Behavior Probabilities}
\begin{figure}[th!]
\centering
    \subfloat[An example of smooth probabilities. The trace is malicious and the probabilities of being benign are mostly 0.]{
    	\includegraphics[scale=0.4]{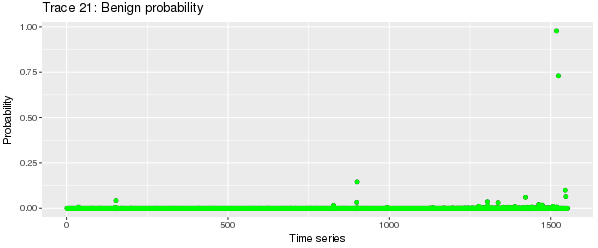}
        \label{fig:ts_prob_good}
	}	
    \\
	\subfloat[A malicious trace with anomalous behavior. We discover that towards the tail of execution, a system function is called in a loop, causing the low-level model to misclassify the time period as benign.]{   
    	\includegraphics[scale=0.4]{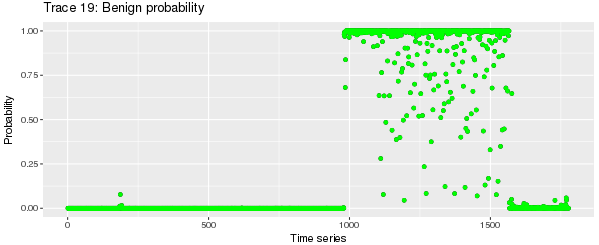}
	\label{fig:ts_prob_bad}
    }
    \caption{Time series behavior probabilities.}
	\end{figure}

The aggregated probability produced by the ensemble model enables malware detection with high confidence.
For the traces with high ensemble probability, the time series of behavioral probabilities are relatively smooth. Refer to Figure ~\ref{fig:ts_prob_good} as an example. We, however, examine a malicious test trace which has a 0.62 probability of being malicious. Here, the decision confidence is lower compared to other test cases. Figure~\ref{fig:ts_prob_bad} shows the execution time-series and reveals that the behaviors towards 20\% from the tail were classified as benign. From manual analysis, we find that the attack invokes a lot of $strcmp()$ function calls in a loop. Since $strcmp()$ is a commonly used function, the malicious execution segments are misclassified as benign causing a lower detection confidence. 

\section{Related Work}
\label{sec:related}
 

There has been little work so far in developing robust deep learning malware classifiers. Existing deep learning approaches use static or dynamic features to construct neural networks for malware detection or classification. The static features are obtained by analyzing an executable file. Neural network malware classifiers have been trained on static features including byte/entropy histogram, imports, raw header bytes and other metadata of Portable Executable files or several million byte long executable files~\cite{Saxe2015}\cite{Raff2017}\cite{RaffSylvester2017}. Despite this success, Grosse et al. showed that such malware detectors can be bypassed by adversarially crafted malware samples~\cite{Grosse2016}.

Dynamic analysis approaches are more robust to obfuscation. Dahl et al. proposed a multi-class malware neural network classifier trained on random projections of features including API call tri-grams, API parameter, and file strings~\cite{Dahl2013}. Pascanu et al. trained recurrent neural nets on API call information for both malware detection and classification~\cite{Pascanu15}. 
Kolosnjaji et al. combine convolutional and recurrent neural networks to optimize malware classification~\cite{Kolosnjaji2016}. API call based approaches require significant domain knowledge and feature engineering. More importantly, an attacker can effectively bypass sequential API feature based malware detectors by generating sequential adversarial examples~\cite{Hu2017}\cite{Rosenberg2017}.

Classical machine learning algorithms have been applied on static binaries, converted to images, for multi-family malware classification~\cite{nataraj2011malware}\cite{nataraj2011comparative}. 
\section{Conclusion and Future Work}\label{sec:conclusion}
HeNet is the first deep learning approach to classify Intel\textsuperscript{\textregistered} PT generated fine-grained control flow traces for malware detection. HeNet is portable across various software systems due to using hardware execution trace based features. HeNet achieves high accuracy and faster training as it builds application behavior model by using transfer learning on a time-series of images, generated from control flow trace. The experimental results show that HeNet detects ROP attacks agianst Adobe\textsuperscript{\textregistered} Reader with 100\% accuracy and 0\% false positive. 

This work is a step towards deep learning analysis of application execution trace for robust malware detection. 
We discuss below our future work in this direction.
 
\subsection{Ensemble Behavior Model}
\begin{figure}
\centering
\includegraphics[scale=0.34]
{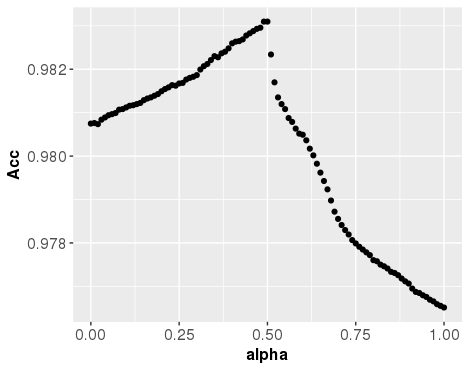}
\includegraphics[scale=0.34]
{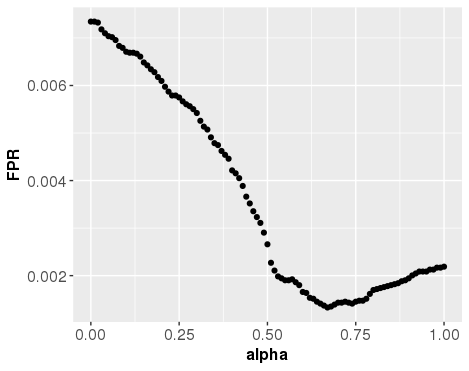}
\caption{Accuracy and FPR for various convex combination parameter $\alpha$. Optimal accuracy and FPR of HeNet and low-res model is at $\alpha=0$ and $\alpha = 1$, respectively.
}
\label{fig:alpha_plot}
\end{figure}

To achieve high accuracy and faster training, the HeNet low-level behavior model is trained via transfer learning on high resolution trace images. We refer to the model as \textit{high-resolution model}. We investigate an ensemble of \textit{high-resolution model} and \textit{low-resolution model}, a model trained from scratch on low resolution images, for further performance improvement. We train ResNet from scratch with 20 layers for three-channel images of size 28$\times$28, learning rate of 0.05, and batch size of 100. We choose the model at 20\textsuperscript{th} epoch. Note that HeNet's behavior model converges within 10 epochs. We apply convex combination on the \textit{low-resolution} and \textit{high-resolution model} so that the new model is $M_{e} = \alpha M_l + (1-\alpha)M_h$, where $\alpha \in [0,1]$ is selected to optimize metrics including accuracy, false positive rate, sensitivity, and area under the curve~\cite{chen2013refinement}.
In Figure \ref{fig:alpha_plot}, we consider an optimal convex combination parameter $\alpha \in [0.48,0.68]$ for both desired accuracy and FPR. Selecting $\alpha = 0.5$, model $M_{e}$ has $0.9831$ accuracy, higher than that of $M_l$ and $0.0029$ FPR, lower than that of $M_h$. We plan to further investigate this model. 

\begin{table} 
\centering
  \caption{Performance of ensemble behavior model}
  \begin{tabular}{||l|l|l|l||}
  \hline
    Behavior model &  Accuracy & FPR & Resolution \\ \hline  
    Ensemble low-level model $M_e$ & 0.983 & 0.0029 & High-low res \\
     HeNet low-level model $M_h$ & 0.981 & 0.0073 & 224$\times$224$\times$3 \\
     ResNet Low-resolution model $M_l$ & 0.976   &  0.0022& 28$\times$28$\times$3\\
     \hline
\end{tabular}
\label{tab:convex_compare}
\vspace{-0.2in}
\end{table}

\subsection{Adversarial Machine Learning}
Studies show that small adversarial purturbations, even unnoticable by human eyes, may lead to misclassification by deep learning image classifiers~\cite{goodfellow2014explaining}\cite{moosavi2016universal}\cite{carlini2016towards}. As we assume end-to-end trace security, to fool the HeNet behavior classification, the attacker goal is to inject malicious control flow transfers at runtime in a way that also keeps the application operational. Determinstic control flow integrity methods are shown to be vulnerable due to the weakness of control flow graph and security policies~\cite{Carlini15}\cite{Carlini14}. HeNet learns the legitimate control flow behavior of an application from training on dynamic traces. As we use both conditional and indirect branches, the behavior model is fine-grained. Hence, it becomes a difficult challenge for the attacker to effectively manipulate the control flow. We, however, will investigate whether an attacker can bypass the HeNet ensemble model exploiting threshold based decision. We will also study adversarial perturbations to exploit HeNet under certain constraints. 


\subsection{Different Data Representations}
Beyond pixel conversion, TNT/TIP packets may be utilized to extract other useful features. We plan to investigate indirect branch target addresses to build adversary resilient deep learning malware classifier.   
\section*{Acknowledgment}
We would like to thank David Durham at Intel Labs for his insightful discussion and valuable feedback.

\bibliographystyle{IEEEtran}
\bibliography{Henet,references}



%



\end{document}